\documentclass[12pt]{iopart}

 \usepackage{cite}
\usepackage{graphicx}
\usepackage{float}
\usepackage{latexsym}
\usepackage{stmaryrd}
\usepackage{iopams}  

\usepackage[ngerman, english]{babel} 

\begin{document}

\title[]{Complex networks embedded in space: Dimension  and
scaling relations between mass, topological distance and Euclidean distance
}

\author{Thorsten Emmerich$^1$, Armin Bunde$^1$, Shlomo Havlin$^2$, Li Guanlian$^3$ and Li Daqing$^4$}
\address{$^1$Institut f\"ur Theoretische Physik, Justus-Liebig-Universit\"at Giessen, 35392 Giessen,Germany}
\address{$^2$Minerva Center and Department of Physics, Bar-Ilan University, Ramat-Gan 52900, Israel}
\address{$^3$Center for Polymer Studies, Department of Physics, Boston
  University, Boston 02215,USA}
\address{$^4$School of Reliability and Systems Engineering, Beihang University, Beijing, 100191, China}

 \ead{thorsten.emmerich@theo.physik.uni-giessen.de}

\begin{abstract}
Many real networks are embedded in space, where in some of them the links length decay as a power law
distribution with distance. Indications that such systems can be characterized by the concept of
dimension were found recently. Here, we present further 
support for this claim, based on extensive numerical simulations for model networks embedded on lattices of 
dimensions $d_e=1$ and $d_e=2$.
 We evaluate the dimension $d$ from the power law scaling of (a) the
mass of the network with the Euclidean radius $r$ and (b) the probability of
return to the origin with the  distance $r$ travelled by the random
walker. Both approaches yield the same dimension. For networks with $\delta <
d_e$, $d$ is infinity, while for $\delta > 2d_e$, $d$ obtains the value of the
embedding dimension $d_e$. In the intermediate regime of interest $d_e \leq
\delta < 2 d_e$, our numerical results suggest that $d$ decreases continously from $d = \infty$ to $d_e$, with $d
- d_e \sim (\delta - d_e)^{-1}$ for $\delta$ close to $d_e$. Finally, we
discuss the scaling of the mass $M$ and the Euclidean distance $r$ with the
topological distance $\ell$ (minimum number of links between two sites in the network). 
Our results suggest that in the intermediate regime $d_e \leq
\delta < 2 d_e$, $M(\ell)$ and $r(\ell)$ do not increase with $\ell$ as a
power law but with a stretched exponential, $M(\ell) \sim \exp [A \ell^{\delta'
    (2 - \delta')}]$ and $r(\ell) \sim \exp [B \ell^{\delta' (2 - \delta')}]$,
where $\delta' = \delta/d_e$. The parameters $A$ and $B$ are related to $d$ by
$d = A/B$, such that $M(\ell) \sim r(\ell)^d$. For $\delta < d_e$, $M$
increases exponentially with $\ell$, as known for $\delta=0$, while $r$ is constant and independent of $\ell$. For 
$\delta \geq 2d_e$,
we find power law scaling, $M(\ell) \sim \ell^{d_\ell}$ and $r(\ell) \sim
\ell^{1/d_{min}}$, with $d_\ell \cdot d_{min} = d$. For networks embedded in
$d_e =1$, we find the expected result, $d_{\ell} = d_{min} = 1$, while for networks
embedded in $d_e = 2$ we find surprisingly,  that although $d=2$, $d_{\ell} > 2$ and $d_{min} < 1$, in contrast to
regular lattices.
\end{abstract}

\submitto{\NJP}

\maketitle
\section{Introduction}
It has been realized in the last decades that a large number of  complex systems are structured in
the form of networks. The structures can be
man-made like the World Wide Web and transportation or
power grid networks or natural like protein and neural networks
\cite{Albert1999,Albert2002,Watts1998, Watts1999,Newman2002,Milo2002,Dorogovtsev,Satorras2004:Cambridge,Gallos2005,Bollobas2001,
Barrat2008,Brockmann2006,Cohen200,Yanqing2011,Newmann2010:Oxford,Csanyi, Gastner Cohen2010,Cohen2003:PhyRevLett}. 
When studying the properties of these networks it is usually assumed
that spatial constraints can be neglected. This
assumption is certainly correct for networks like the World Wide Web (WWW) or the citation network
where the real (Euclidean) distance does not play any
role, but it may not be justified in networks where
the Euclidean distance matters \cite{Barth2011:PhysRep}. Typical examples of such networks include the Internet
\cite{Albert2002,Satorras2004:Cambridge}, airline networks 
\cite{Barrat2004,Bianconi2009}, wireless
communication networks \cite{Hao2009}, and social networks (like friendship and author networks)
\cite{Nowell2005,Lambiotte2008}, which are all embedded
in two-dimensional space (surface of the
earth), as
well as protein and neural networks \cite{Jeong2001}, which are embedded in three dimensions.
\\
\indent To model these networks, two network classes are of particular interest:
Erd$\ddot{o}$s-R$\grave{e}$nyi (ER) graphs
\cite{ER1960,ER1959} and Barabasi-Albert (BA) scale free
networks \cite{Barabasi1999}. In
ER-networks, the distribution of the number $k$ of links
per node (degree-distribution) is Poissonian with a pronounced maximum at a certain $k$-value, such
that nearly each node
is linked to the same number of nodes. In BA
networks, the distribution follows a power law $P(k)
\sim k^{-\alpha}$, with $\alpha$ typically between 2 and 3. Here we focus on ER-type networks
embedded in one- and two-dimensional space. We actually
use a degree distribution that is close to a delta
function (as the case in simple lattices). We found that the results are the same for both
kinds of distributions. We
follow Refs. \cite{daqingPerc2011,kosmidis2008,daqingDim2011} and assume that nodes
are connected to each other with a probability $p(r) \sim r^{-\delta}$, where $r$ is the Euclidean
distance between the nodes. The choice of a power
law for the distance distribution is supported from
findings in the Internet, airline networks, human travel networks and other social networks 
\cite{Bianconi2009,Lambiotte2008,Goldberg2009}. 
Our
model of embedding links of length $r$, chosen from  Eq. ({\ref{prD}}), in a $d_e$-
 dimensional lattice can be regarded as a generalization of the known Watts
 Strogatz (WS) model \cite{Watts1998, Watts1999}. In the WS model links of any possible lengths with the
 same probability are added in the lattice system which corresponds to the
 case $\delta=0$ of Eq. ({\ref{prD}}). Other methods for embedding networks in Euclidean space have been
proposed in \cite{Rozenfeld2002, Manna2003, Sokolov1, Sokolov2}.
\\
 \indent It has recently been shown that spatial constraints are important and may alter the
 dimension and therefore the  topological properties of the networks (likethe dependence of the mean
 topological distance on the system size) as well as their robustness \cite{daqingPerc2011,kosmidis2008}.
Here we are interested in studying how in these model networks the spatial constraints
quantified by the distance exponent $\delta$ modify the scaling
relations between mass (number of nodes), Euclidean
distance $r$ and topological distance $\ell$. Our earlier study on ER networks embedded in a square
lattice (with  dimension
 $d_e=2$),  indicate that by varying    the  exponent $\delta$ one can
 actually change continuously the {\it dimension} $d$ of the network,  from $d=\infty$ for  
 $\delta<2 $ to $d=2$ for $\delta>4$  \cite{daqingDim2011}. In the present manuscript we present further extensive
numerical simulations for $d_e=2$ that support this claim as well as
simulations in linear chains ($d_e=1$) that suggest analogous conclusions. In $d_e=1$ we find
that for $\delta<1$  the system behaves like an infinite dimensional network
(as the original  ER-network). When continuously increasing $\delta$ the dimension becomes
 finite for $\delta>1$ and approaches $d=1$ for $\delta>2$.  Since the dimension of a system
plays a critical role in many physical phenomena like diffusion,
percolation and phase transition phenomena, our results are important for
understanding and characterizing the properties of real world networks. \\
\indent Our manuscript is organized as follows. In Section \ref{ChapDistances}, we discuss the
characteristic distances in the spatially constrained networks.
In Section \ref{ChapGenNet}  we describe the  method to generate the spatial network models. In
Section \ref{ChapMR} we present our numerical 
results for the dimension $d$, for networks embedded in linear chains and in square lattices, that
we obtain from the scaling relation of the mass
$M $ and the distance $r$. In Section \ref{ChapP0} we present our numerical 
results for the dimension $d$, that we obtain from the scaling relation of the
probability of return to the origin $P_0$ of a diffusing particle and its distance $r$.
In Section \ref{ChapMlRl}, we discuss the scaling of the mass $M$ and the Euclidean distance $r$ with the
topological distance $\ell$. The conclusions in Section \ref{summary} summarize our main results.
\begin{figure}[tbh!]
  \centering
  \includegraphics[angle=0, scale=0.35]{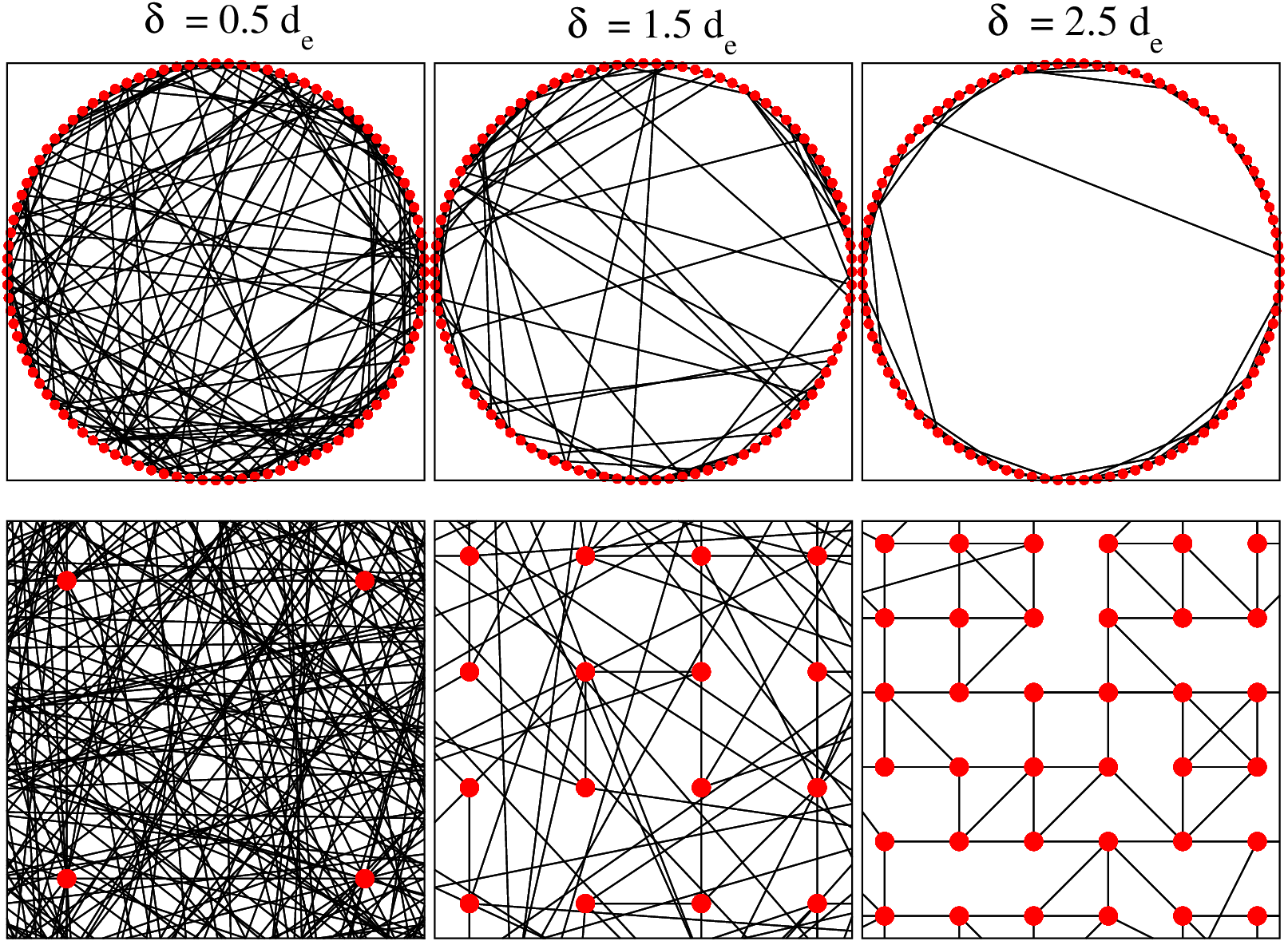}
  \caption{Illustration of ER networks embedded in linear chains (top) and 
    square lattices  (bottom), for various distance exponents $\delta$.}
  \label{network}
\end{figure}
\section{Characteristic distances}
\label{ChapDistances}
\noindent
First we estimate how the characteristic distances, in a network of $L^{d_e}$ nodes, depend on its
linear size $L$, on $\delta$ and on the embedding dimension $d_e$. 
We normalize the distance distribution $p(r)$ such that $\int^L_1  dr \, r^{d_e-1}p(r)=1$, which yields 
\begin{equation}
 p(r) = \left\{
    \begin{array}{ll}
     (d_e-\delta)L^{-(d_e-\delta)} \, r^{-\delta}       & ,     \delta < d_e   \\[3mm]
     (\delta - d_e) \, r^{-\delta} &, \delta > d_e.
    \end{array}
\right.
\label{prD}
\end{equation}
\noindent
From $p(r)$ we obtain $\overline{r^n} = \int^L_1 dr \, r^{d_e-1} \, r^n \,p(r)$ 
and the related length scales $\bar{r}_n \equiv (\overline{r^n})^{1/n}$.  
The maximum distance $r_{max}$ is determined by 
$L^{d_e} \int^L_{r_{max}}dr \, r^{d_e-1}p(r) \simeq 1$. The results for $\overline{r^n}$ and $r_{max}$ are
\begin{equation}
 \overline{r^n} = \left\{
    \begin{array}{ll}
     \frac{d_e - \delta}{d_e +n - \delta}\,L^n       & ,     \delta < d_e   \\[3mm]
     L^n/\textrm{ln}(L)                            & , \delta = d_e       \\[3mm]
     \frac{\delta - d_e}{d_e +n - \delta}\,L^{d_e+n-\delta} & , d_e < \delta < d_e+n \\[3mm]
     n \, \textrm{ln}(L) & , \delta = d_e+n \\[3mm]
      \frac{d_e - \delta}{d_e +n -\delta} & , \delta > d_e +n
    \end{array}
\right.
\label{rbar}
\end{equation}
\noindent
and
\begin{equation}
 r_{max} \simeq \left\{
    \begin{array}{ll}
     L       & ,     \delta < 2 d_e   \\[3mm]
     L^{d_e/(\delta-d_e)} &, \delta \ge 2d_e.
    \end{array}
\right.
\label{rmax}
\end{equation}
\noindent
Accordingly, for $\delta < d_e$ all length scales ($\bar{r}_n$ and $r_{max}$) are proporional to $L$, the 
spatial constraints are weak and the system can be regarded as an infinite dimensional system.  On
the other hand, for $\delta > 2 d_e$, $\bar{r}_n
/L$ and 
$r_{max}/L$ tend to zero in the asymptotic limit. In this case, we expect that the physical 
properties of the network are close to those of regular lattices of dimension $d_e$. However,  large
finite size effects are expected for $\delta$
close to $2 d_e$ where 
$r_{max}/L$ decays only very slowly to zero. In the intermediate $\delta$-regime $d_e \leq \delta < 2d_e$,
$r_{max}$ scales as $L$, while $\bar{r}_n / L$  tends to zero in the asymptotic limit. 
In this regime our simulation results (Chap. \ref{ChapMR}) suggest intermediate behavior represented
by a dimension between $d_e$ and
infinity that changes with $\delta$.
\section{Generation of the networks}
\label{ChapGenNet}
\noindent The nodes of the network are located at the sites of a $d_e$-dimensional regular lattice,
in our case a linear chain of length $L$ ($d_e=1$) or a square lattice of size $L \times L$
($d_e=2$). We assign to each node a fixed number $k$ of links (in most cases, $k = 4$). 
Actually this  network is a
random regular (RR) network since all nodes have the
same degree. It is expected (and we have also verified it numerically) that  both networks, ER and RR
with the same spatial constraints, are in the same universality class. \\ 
\indent To generate the spatially embedded networks, we use the following iterative algorithm: (i)
We pick a node $i$ randomly and choose, for one of its available $k_i$ links, a distance $r$ $
(1\leq r \leq L)$ from the given probability distribution $p(r)$,  Eq. (\ref{prD}). 
It is easy to see that the distance $r$ can be obtained from random numbers $0 < u \leq 1$ chosen
from the uniform distribution, by
\begin{equation}
  r = \left\{
    \begin{array}{ll}
      [1 - u (1 - L^{d_e - \delta})]^{{1}\over{d_e - \delta}}     & ,  \delta\ne d_e   \\[2mm]
     L^u      & ,    \delta=d_e.
    \end{array}
\right.
\end{equation}   
\begin{figure}[tbh!]
  \centering
  \includegraphics[angle=0, scale=0.4]{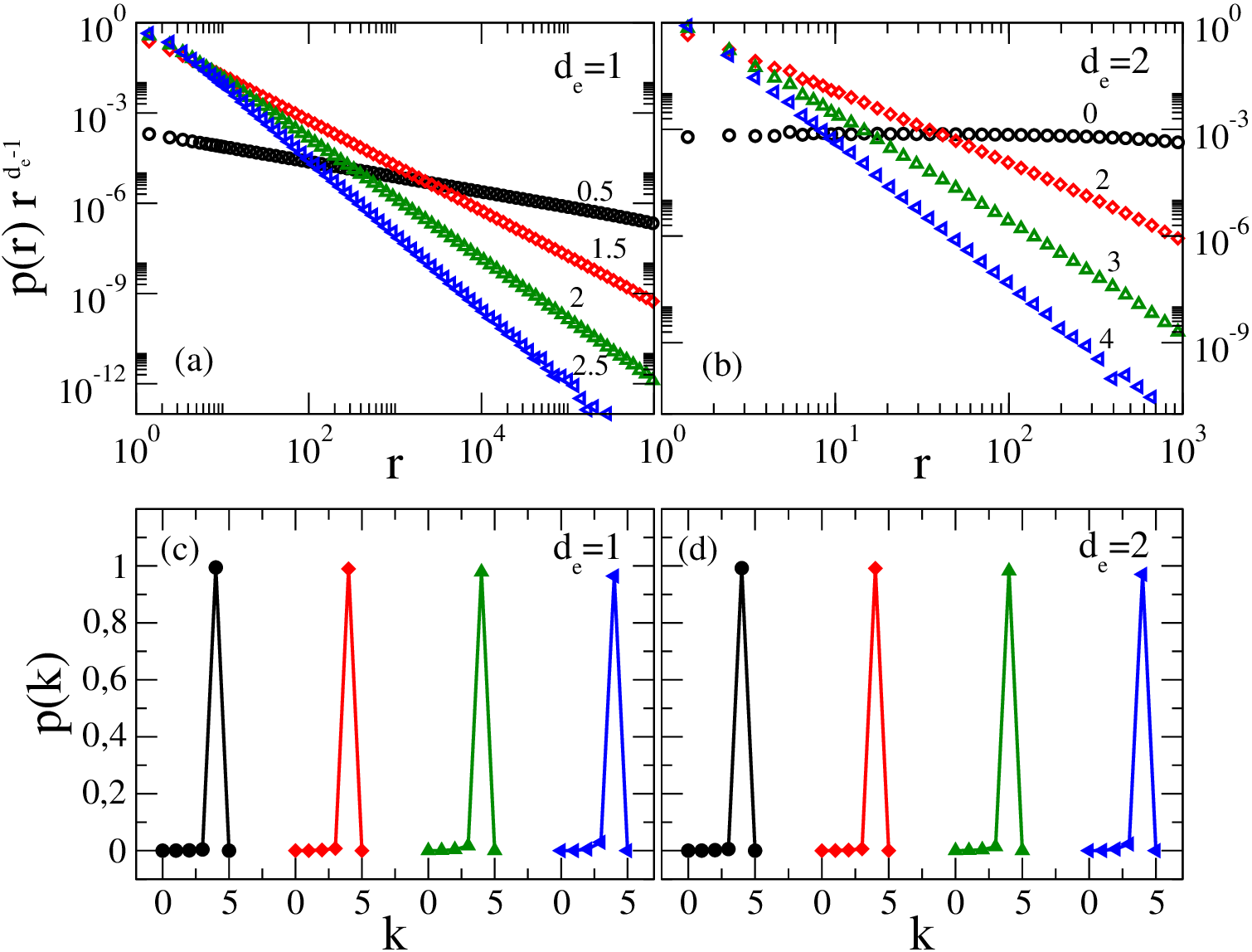}
  \caption{The distance distribution $p(r)\, r^{d_e-1}$  for ER-networks embedded (a) in linear
    chains where $d_e=1$ and (b) in square lattices where $d_e=2$, when 
    $\delta=0.5d_e$(circle), $1.5d_e$(diamond), $2d_e$(triangle up), $2.5d_e$(triangle left)
    and $k=4$. The numbers denote the slopes of $P(r)r^{d_e-1}$, which are identical to the anticipated ones.
    For the same set of parameter as in (a) and (b), the panels (c) and (d) show the 
    degree distribution $p(k)$, which is $\cong 1$ for $k=4$ and $\cong0$ otherwise.
  }
  \label{PRK} 
\end{figure}
\noindent
 (ii) We
consider all $N_r$ nodes between distance $r - \Delta r$ and $r$ from node $i$, that are not yet
connected to node $i$. Without loss of generality, we choose $\Delta r = 1$ for the linear chain and
$\Delta r = 0.4$ for the
square lattice.
(iii) We pick randomly one of these nodes $j$. If node $j$ has at least one available link, we
connect it with node $i$. If not, we do not connect it. Then we return to (i) and proceed with
another randomly chosen node. At
each step of the process, either 2 or zero links are added. For generating the network, we have
typically performed $10^3 \cdot L^{d_e}$ trials. Due to the generation process, the nodes of the
final network do not all have exactly the same degree, but the degree follows a narrow distribution with a
mean $\bar{k}$ slightly below $k =4$. Figure \ref{network} illustrates the ER networks embedded in
$d_e$=1 and $d_e$=2 for $\delta= 0.5 d_e ,  \, 1.5 d_e \,\,\textrm{and} \,\, 2.5d_e$. Figure \ref{PRK} shows
the actual narrow degree distribution as well as $p(r)$ obtained in the simulations.
\section{The dimension of the networks}
\label{ChapMR}
\noindent For determining the dimensions of the spatially embedded networks, we follow the method
developed by Daqing et al \cite{daqingDim2011}. 
We use the fact that the mass $M$ (number of nodes) of an object within an hypersphere of radius $r$
scales with $r$ as
\begin{equation}
M \sim r^d
\label{MR}
\end{equation}
\noindent where the exponent $d$ represents the dimension of the network. When using this relation
without taking into account the way the nodes are linked, one trivially and erroneously finds that the dimension of
the network is identical to the dimension $d_e$ of the embedding space.\\
\begin{figure}[htb!]
  \centering
  \includegraphics[angle=0, scale=0.35]{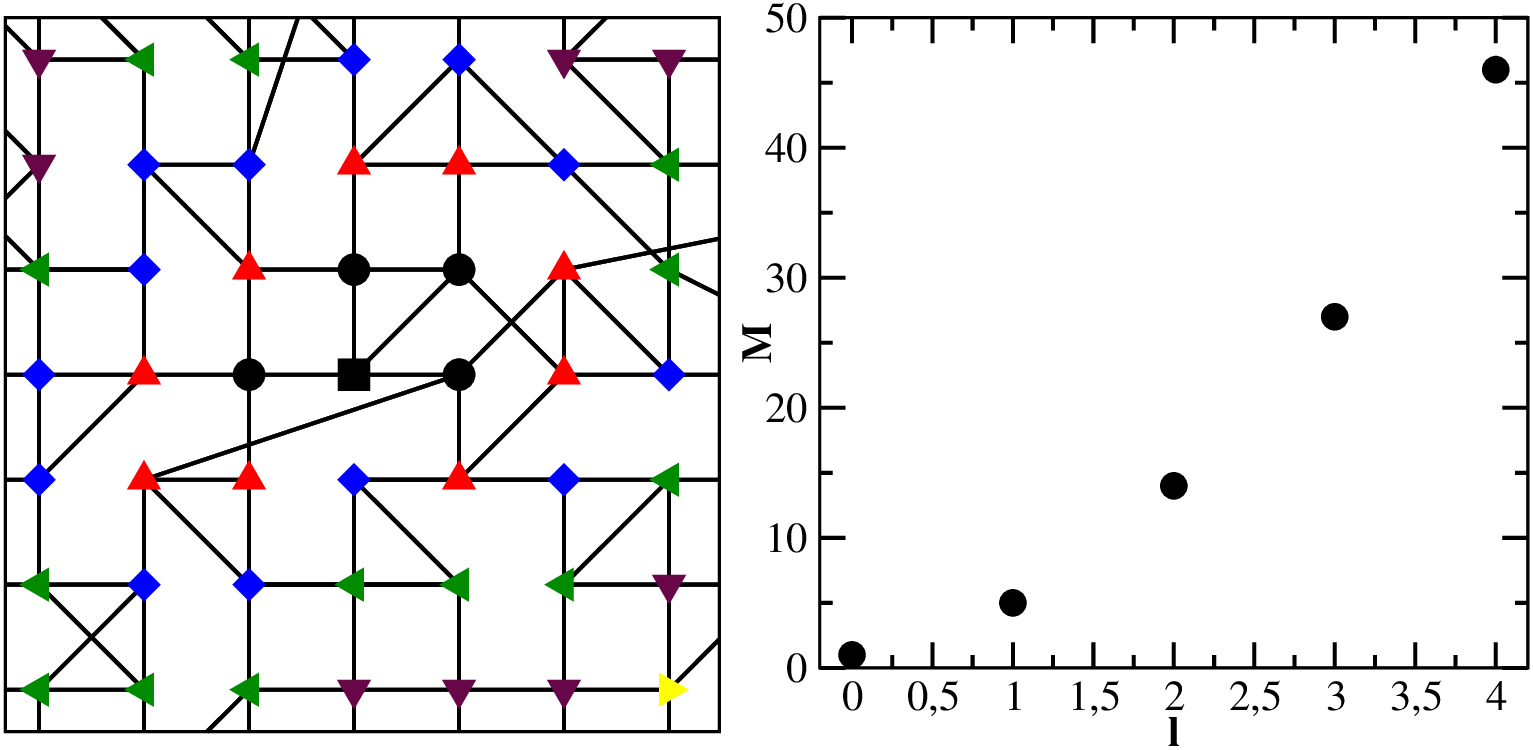}
  \caption{
    Illustration of the shells 
    $S(0)$ (square), $S(1)$ (circle), $S(2)$ (triangle up), $S(3)$ (diamond),
    $S(4)$ (triangle left),   $S(5)$ (triangle down) and  $S(6)$ (triangle right)
    for ER networks embedded in a square lattice with $k=4$ (left panel), and the mass $M$ 
    as function of $l$ within this shells (right panel).
  }
  \label{shel} 
\end{figure}
\indent To properly take into account the connectivity, when considering the dimension of the network, we
proceed as follows (see
\begin{figure}[htb!]
\centering
  \includegraphics[angle=0, scale=0.45]{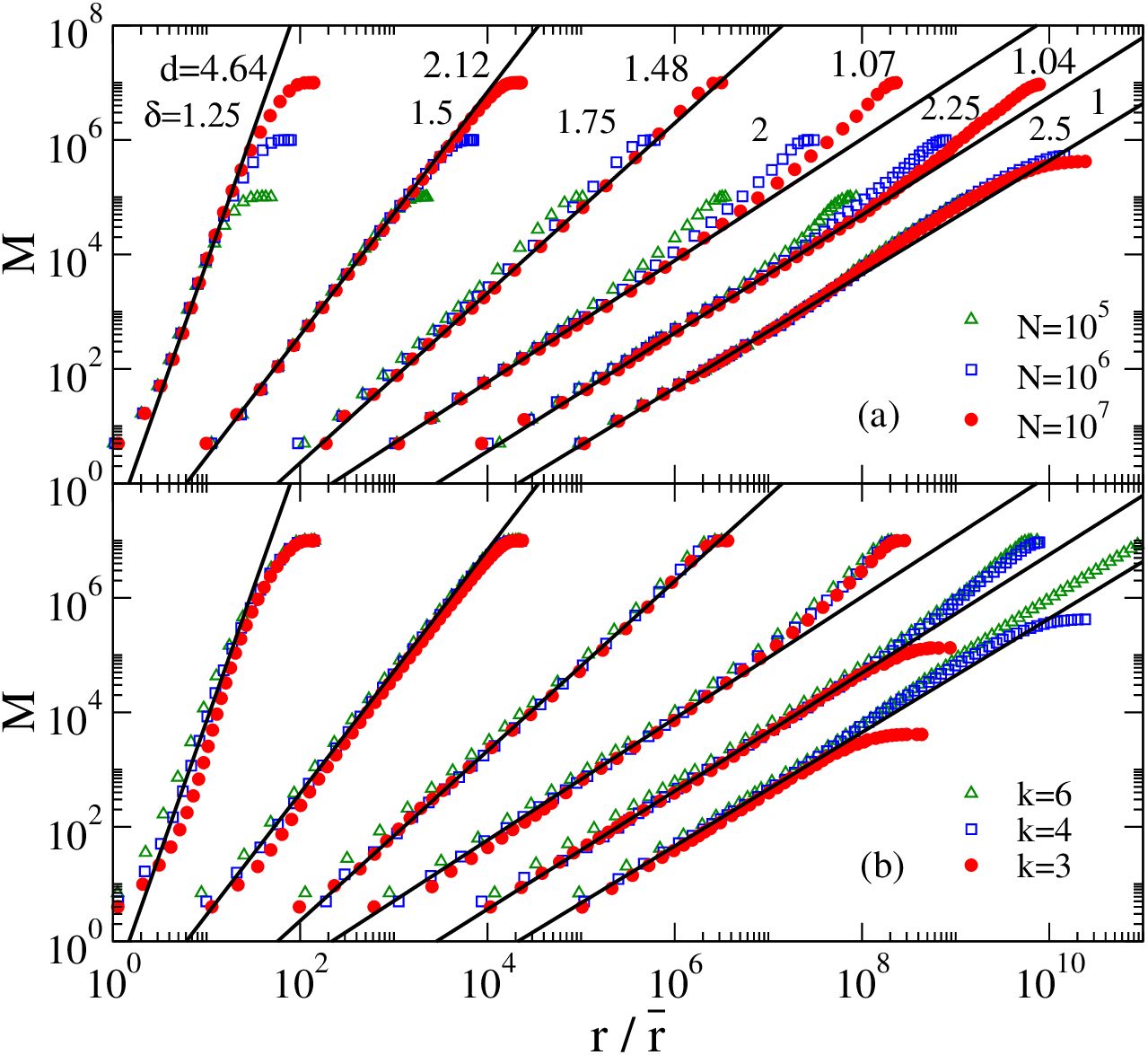}
\caption{(a) The mass $M$ as function of the relative distance $r/ \bar{r}$ for ER networks embedded in
  linear chains with $k=4$, for the system sizes $N=10^5,\, 10^6 \,\textrm{and}\, 10^7$ with
  $\delta=1.25, \, 1.5, \, 1.75, \, 2, \, 2.25, \, 2.5$ (from left to right). 
The straight lines are best fits to the data that yield the dimension $d$ of the network.
(b) The same as panel (a), but for $N=10^7$ and $k=3, 4 \, \rm{and} \, 6$.
}
 \label{MR_1d} 
\end{figure}
Fig \ref{shel}): We choose a node as origin and determine its nearest
neighbors (referred to as shell 1) and their number $S(1)$, the number of second nearest neighbors $S(2)$, and so
on. Next we measure the mean Euclidean distance $r(\ell)$ of the nodes in shell $\ell$ from the
origin and determine the number of nodes $M(\ell) = \sum^\ell_{i = 1} S(i)$ within shell $\ell$. 
To improve the statistics, we repeat the calculations for many origin nodes and then
average $r(\ell)$ and $M(\ell)$. To reduce finite size effects, we do not choose the origin nodes
randomly in the underlying lattice, but from a region with radius $L/10$ around the central node. From
the scaling relation between the average $M$ and the average $r$, Eq. (\ref{MR}), we determine the
dimension $d$ of the network.\\
\indent Figure \ref{MR_1d} shows the results for networks embedded in linear chains, for distance
exponents $\delta$ between 1.25 $d_e$ and 2.5 $d_e$. In (a), we consider networks with  $k=4$ fixed and different 
system sizes ($N = 10^5$, $10^6$ and $10^7$), while in (b) we consider networks with a fixed size
$N = 10^7$ and various  $k$ values ($k=3,4,6$).  In
both panels, we have plotted $M$ as a
function of $r/\bar{r}$,  where 
$\bar{r} \equiv \bar{r}_1$ is the mean distance, see Eq. (\ref{rbar}).\\
\indent Figure \ref{MR_1d}a  shows
that for $\delta$ in the interesting regime between $d_e$ and $2 d_e$, the curves for different $N$
collapse nicely (For transparency, the curves (except $\delta = 1.25$) have been shifted along the
$x$-axis by a factor of $10,\, 10^2,\, 10^3,\, 10^4$ \,\textrm{and}\, $10^5$). From the slopes of the straight
lines, we obtain the dimension 
$d \cong 4.64$ $(\delta = 1.25)$, $d \cong 2.12$ $(\delta = 1.5)$ and $d \cong 1.48$ $(\delta = 1.75)$.
For $\delta \geq 2$, the data starts to overshoot above some crossover
value that increases with the system size and thus can be regarded as a finite size effect. To
understand the reason for this crossover note that a
node close to the boundary has a considerably higher probability to be linked with nodes closer to
the center of the underlying lattice. As a consequence, for large shell numbers $\ell$, the mean
Euclidean distance of the nodes from the origin node will be underestimated and thus the mass within
large Euclidean distances overestimated. This
effect is most
pronounced in the linear chain, for intermediate $\delta$-values, and gives rise to the
overshooting of $M(r)$ for $\delta$ between 2 and 2.5, where $d \simeq d_e$.
For $\delta = 2.5$ and
$N = 10^7$, the total number of nodes in the spatially constrained network is well below $N$, since
the network is separated into smaller clusters. For larger $k$-values, this effect is less likely to
appear. Figure \ref{MR_1d}b shows that the dimension of the networks does not depend on their
average degree. The $M(r)$ curves collapse for different $k$,
and thus give rise to the same dimensions. This indicates the universality feature of the
dimension.\\ 
\begin{figure}[htb!]
\centering
  \includegraphics[angle=0, scale=0.45]{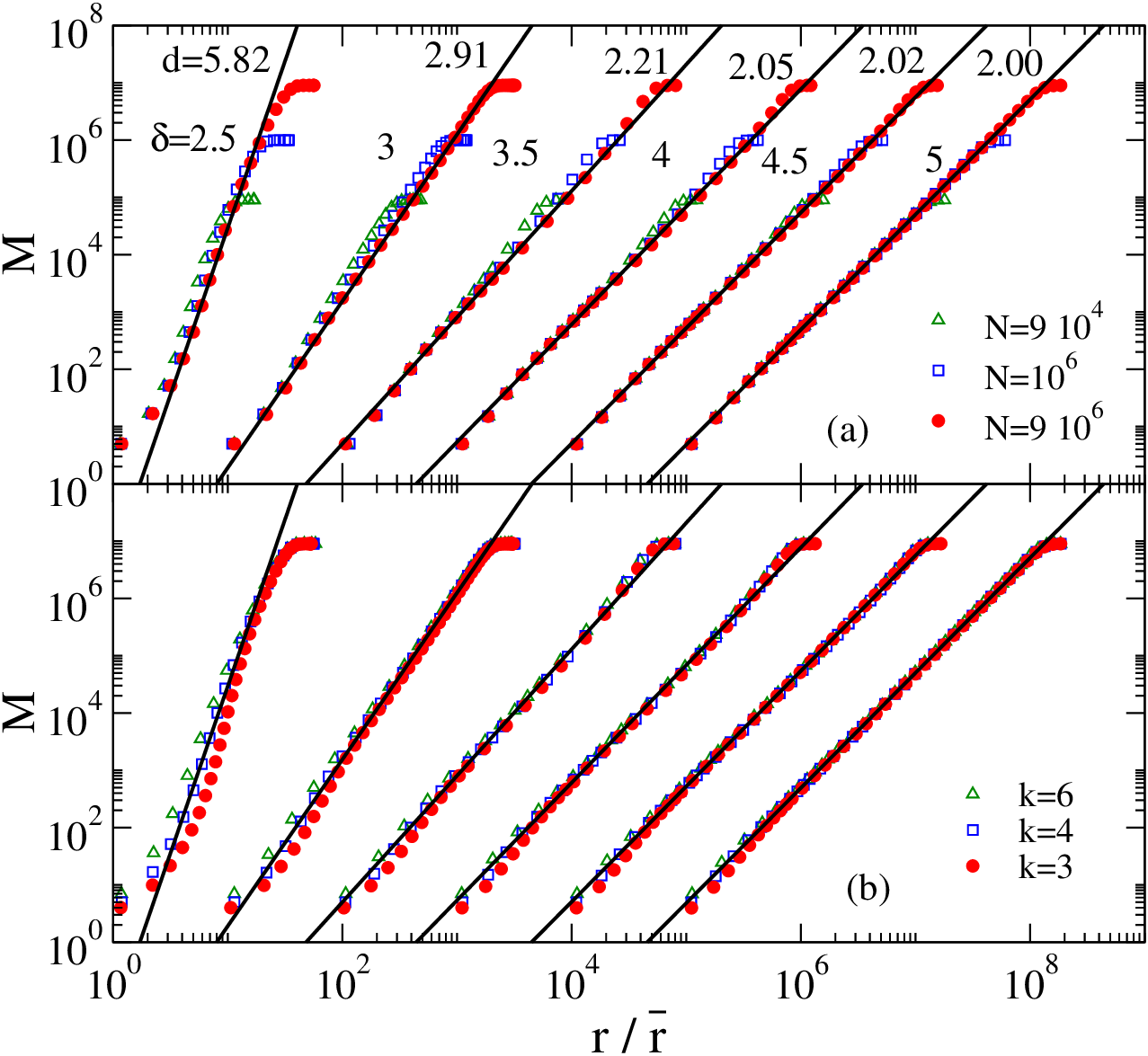}
\caption{The same as Fig \ref{MR_1d}, but for ER networks embedded in a square lattice, the system sizes 
are $N=9 \cdot 10^4,\, 10^6 \,\textrm{and}\, 9 \cdot 10^6$ 
with $\delta=2.5, \,3, \,3.5, \,4, \,4.5, \,5$ (from left to right).}
\label{MR_2d} 
\end{figure}
\indent Figure \ref{MR_2d} shows the corresponding results for networks embedded in square lattices ($d_e=2$),
again for 6 exponents $\delta$ between 1.25 $d_e$ and 2.5 $d_e$, three network sizes ($N=9 \cdot
10^4, 10^6 \, \rm{and} \, 9 \cdot 10^6$), and three $k$ values
($k=3,4,6$). From the slopes of the straight lines we obtain 
$d \cong 5.82$ $(\delta = 2.5)$, $d \cong 2.91$ $(\delta = 3)$, and $d \cong 2.21$ $(\delta = 3.5)$. 
For $\delta$ above 4, $d$ is close to $d_e$, as expected. The figure confirms that the finite size
effects in $d_e= 2$ are considerably less pronounced than in $d_e = 1$, contrary to the intuition,
since the linear size of the underlying embedding
lattice
is considerably higher in $d_e=1$ than in $d_e=2$. As in $d_e=1$, the dimensions are independent of
the mean degree of the networks.\\
\begin{figure}[tbh!]
\centering
  \includegraphics[angle=0, scale=0.35]{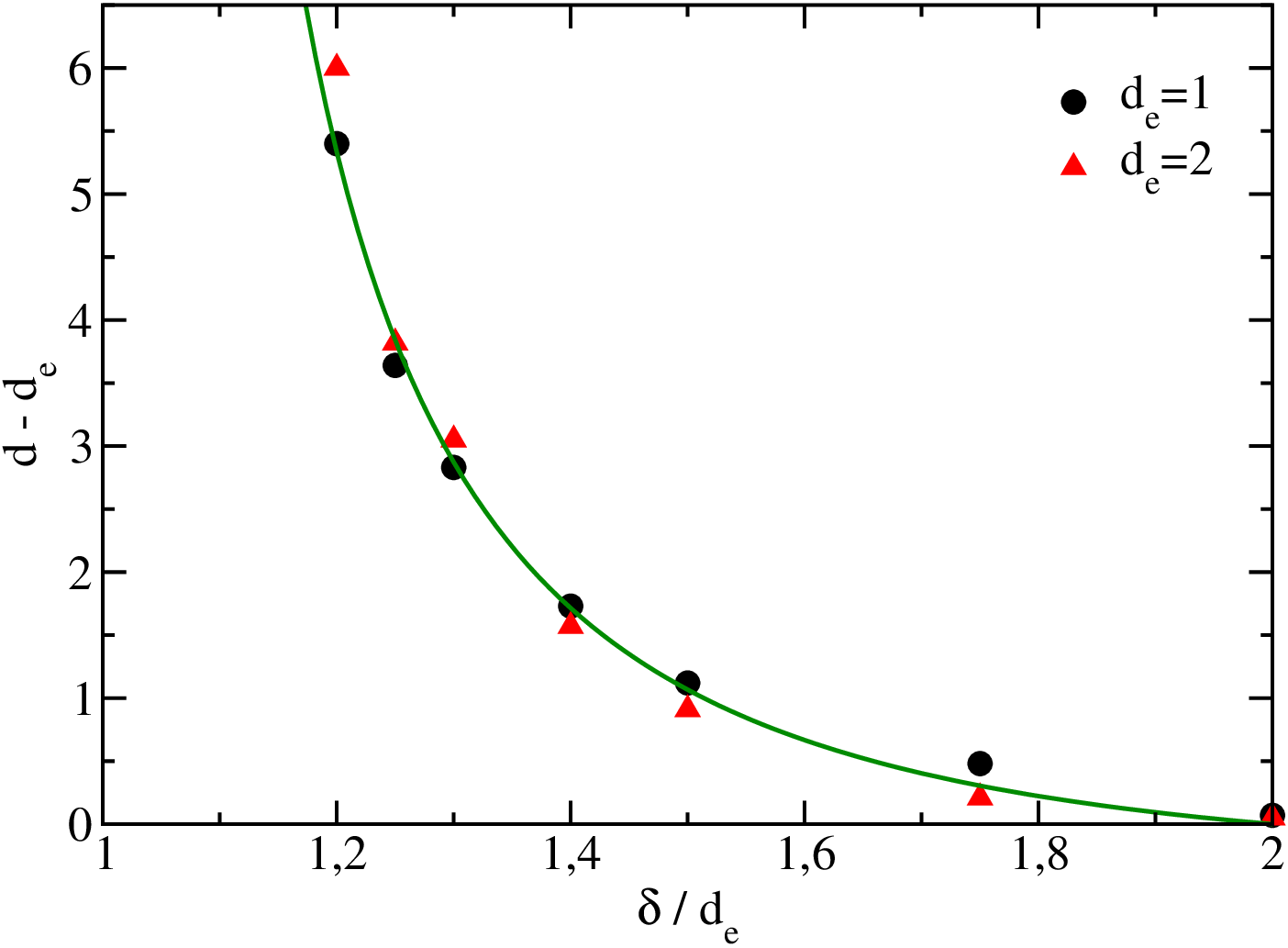}
\caption{The difference between network dimension $d$ and embedding dimension $d_e$  as a function of
  $\delta/d_e$ for $d_e=1$ (circles) and $d_e=2$ (triangles).}.
 \label{dimension} 
\end{figure}
\indent Figure \ref{dimension} summarizes our results for the dimensions of the spatially embedded networks
in the intermediate $\delta$ regime between $d_e$ and $2d_e$, where the dimension is supposed to
bridge the gap between $d=\infty$ for the unconstrained case $\delta$ below $d_e$ and $d=d_e$ for the highly
constrained case $\delta$ above $2d_e$. The figure shows $d - d_e$ as a function of the relative
distance exponent $\delta^\prime = \delta/d_e$  for both
considered lattices. The figure shows that in both cases,  the curves approximately collapse to a single line which
can be represented by 
\begin{equation}
d-d_e = c \frac{2 - \delta^\prime}{\delta^\prime (\delta^\prime -1)}, \quad 1 < \delta^\prime < 2
\label{dmde}
\end{equation}
where $c \cong 1.60$. 
According to Eq. (\ref{dmde}), $d-d_e$ diverges for $\delta^\prime$ approaching the critical 
relative distance exponent $\delta^\prime=1$.
\section{The probability of return to the origin}
\label{ChapP0}
\noindent The network dimension plays an important role also in physical processes such as
diffusion \cite{Weiss1994,havlin2000, Klafter2011}. The probability $P_0 (t)$ that a diffusing particle, after
having traveled $t$ steps, has
returned to the origin, is related to the root mean square displacement $r(t)$ of the particle by
\cite{daqingDim2011,havlin2000,Aleander1982}
\begin{equation}
P_0 (t) \sim r (t)^{- d}.
\label{P0}
\end{equation}
\begin{figure}[tbh!]
\centering
  \includegraphics[angle=0, scale=0.45]{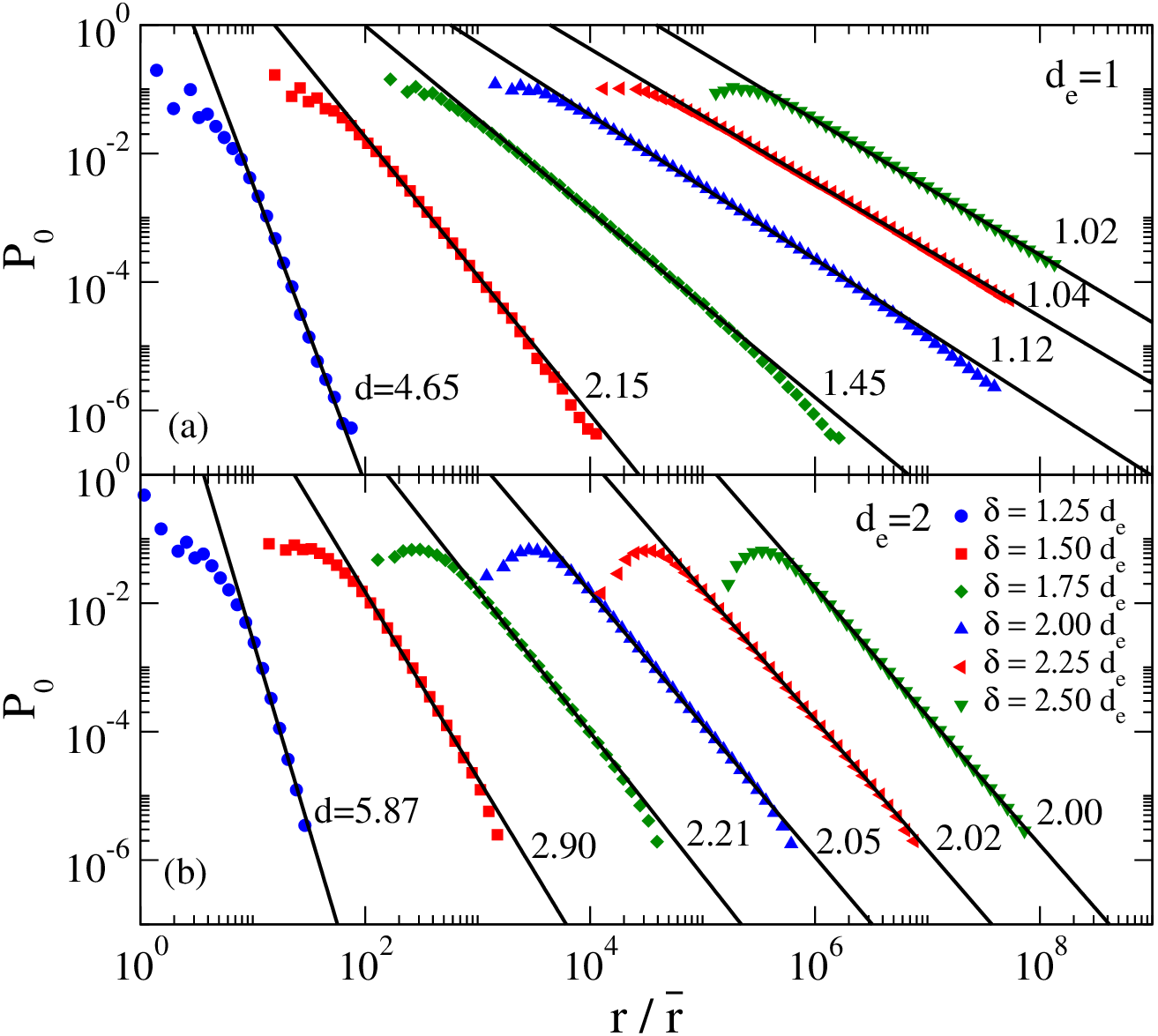}
\caption{(a) The probability $P_0$ that a diffusing particle is at its starting site, after travelling
  an average distance $r$, as a  function of the relative distance $r/ \bar{r}$ for ER networks
  embedded in linear chains with $k=4$, for the system 
size $N=10^7$ with 
$\delta=1.25, \, 1.5, \, 1.75, \, 2, \, 2.25, \,2.5$ (from left to right). 
The straight lines are best fits to the data that yield the dimension $d$ of the network.
(b) The same as panel (a), but for ER networks embedded in a square lattice, the system size is 
$N= 9 \cdot 10^6$ with $\delta=2.5, \,3, \,3.5, \,4, \,4.5, \,5$ (from left to right). 
Note that the values of $d$ obtained here are almost the same as those obtained by direct measurements
in Figs. \ref{MR_1d} and \ref{MR_2d}. 

}
 \label{P0_1d2d}
\end{figure}
\noindent To derive Eq. (\ref{P0}) one assumes that the probability of the particle to be in any
site in the volume $V(t) = [r(t)]^d$ is the same. As a consequence, $P_0 (t) \sim 1/V(t)$, which
leads to Eq. (\ref{P0}).
Figure \ref{P0_1d2d} shows $P_0$ as a function of $r/\bar{r}$
in $d_e = 1$ and 2, for the same $\delta$-values as in Figs. \ref{MR_1d} and \ref{MR_2d}. For
convenience, we show only the results for the largest system size, $N = 10^7$ for $d_e = 1$ and $N =
9 \cdot 10^6$ for $d_e = 2$. To obtain $P_0(t)$, we averaged, for each value of $\delta$, over
$10^4$ diffusing particles and 50 network realizations. From the straight lines in the
double-logarithmic presentations of Figure \ref{P0_1d2d} we obtain the dimension of the
networks, which are listed in the figure. The dimensions obtained in  Figure \ref{P0_1d2d} agree
very well with those obtained by direct measurements  in Figs. \ref{MR_1d}
and \ref{MR_2d}.
\section{The topological dimension  and the dimension  of the shortest path}
\label{ChapMlRl}
\noindent In order to find how $M$ scales with the Euclidean distance $r$, we determined in
Sect. \ref{ChapMR} how $M$ and $r$ scale with the topological length $\ell$, and obtained the dimension $d$
from $M(\ell) \sim r(\ell)^d$. In this section, we discuss explicitely how $M$ and $r$ depend on
$\ell$.\\
\indent It is well known that for regular lattices as well as for fractal structures, $M$ and $r$ scale with
$\ell$ as power laws,
\numparts
\begin{eqnarray}
M(\ell) & \sim & \ell^{d_\ell}\\[1mm]
r(\ell) & \sim & \ell^{1/d_{min}},
\end{eqnarray}
\label{MRL}
\endnumparts
\begin{figure}[tbh!]
\centering
  \includegraphics[angle=0, scale=0.45]{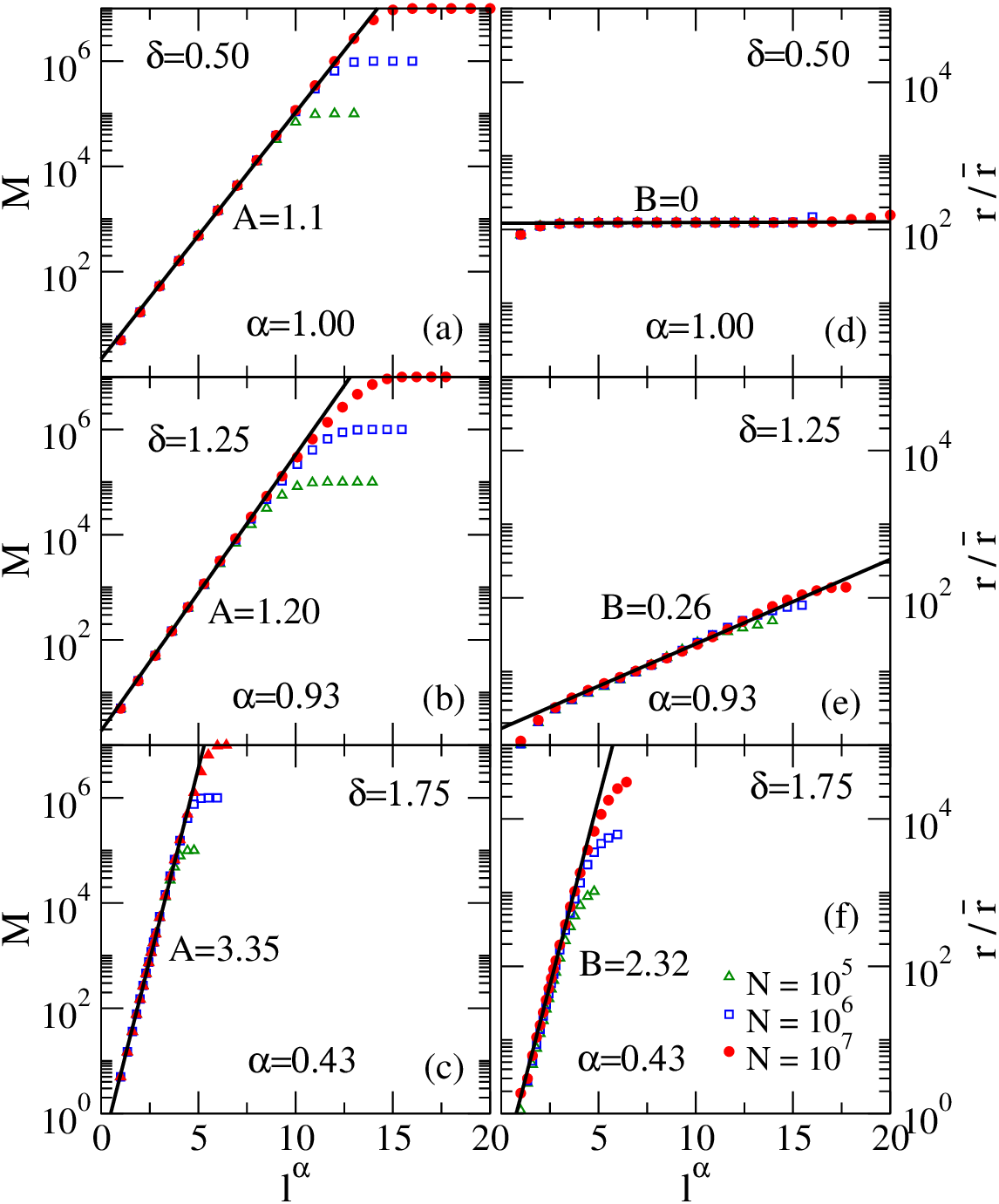}
\caption{The mass $M$ (left column) and the relative distance $r / \bar{r}$ (right column) as function 
of $\ell^{\alpha}$ ($\ell$ is the topological distance and
$\alpha=\delta^{\prime}(2-\delta^{\prime})$) for ER networks embedded in linear chains with $k=4$,
for the system 
sizes $N=10^5,\, 10^6 \,\textrm{and}\, 10^7$ with $\delta=0.5, \, 1.25 \,\textrm{and}\, 1.75$. 
The straight lines are best fits to the data with slopes $A$ and $B$ respectively. 
}
 \label{ML1_1d} 
\end{figure}
\begin{figure}[tbh!]
\centering
  \includegraphics[angle=0, scale=0.45]{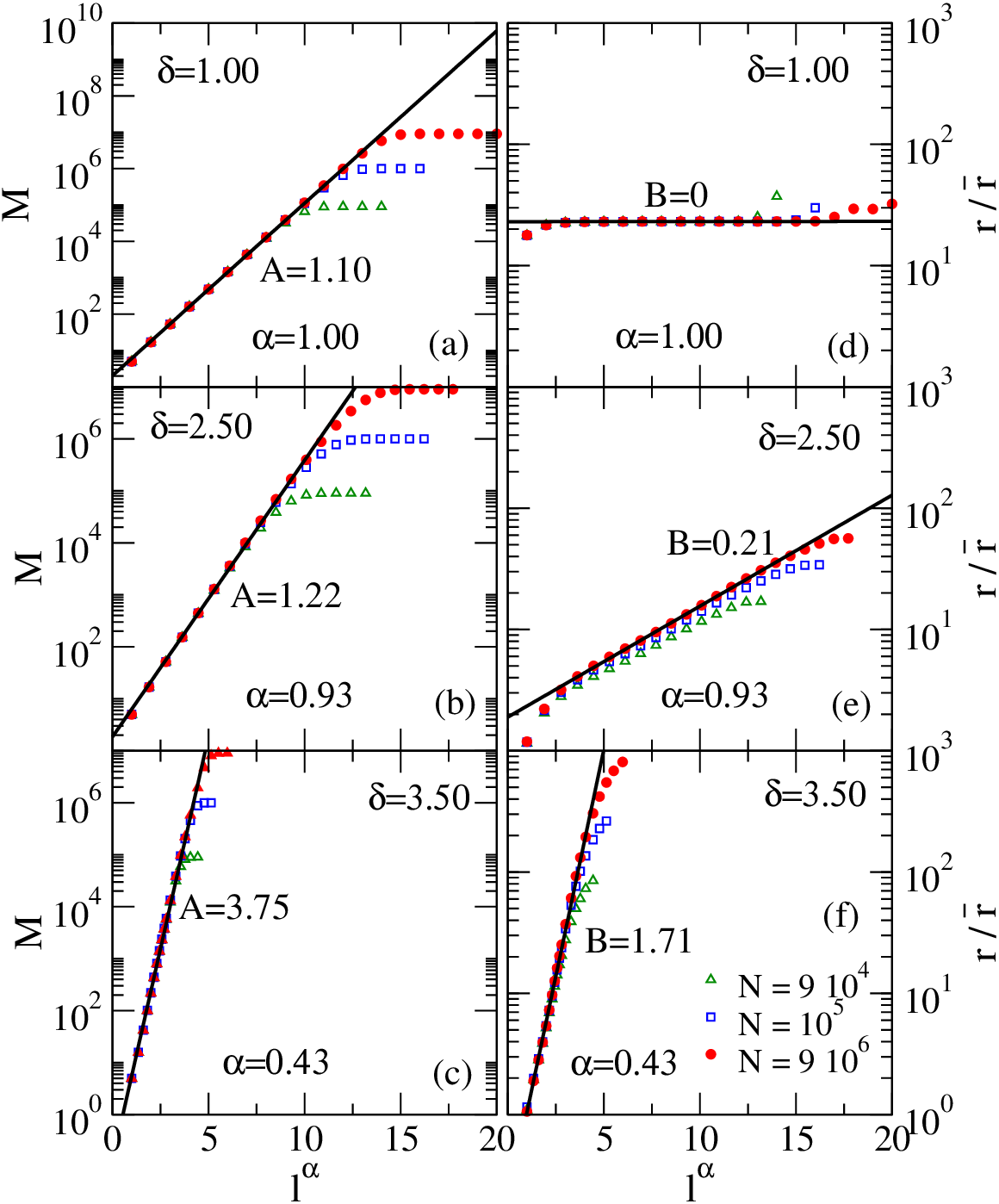}
\caption{The same as Fig \ref{ML1_1d}, but for ER networks embedded in a square lattice, the system sizes 
are $N=9 \cdot 10^4,\, 10^6 \,\textrm{and}\, 9 \cdot 10^6$ 
with $\delta=1.0, \, 2.5 \,\textrm{and}\, 3.5$. 
}
\label{ML1_2d} 
\end{figure}
\begin{figure}[tbh!]
\centering
  \includegraphics[angle=0, scale=0.45]{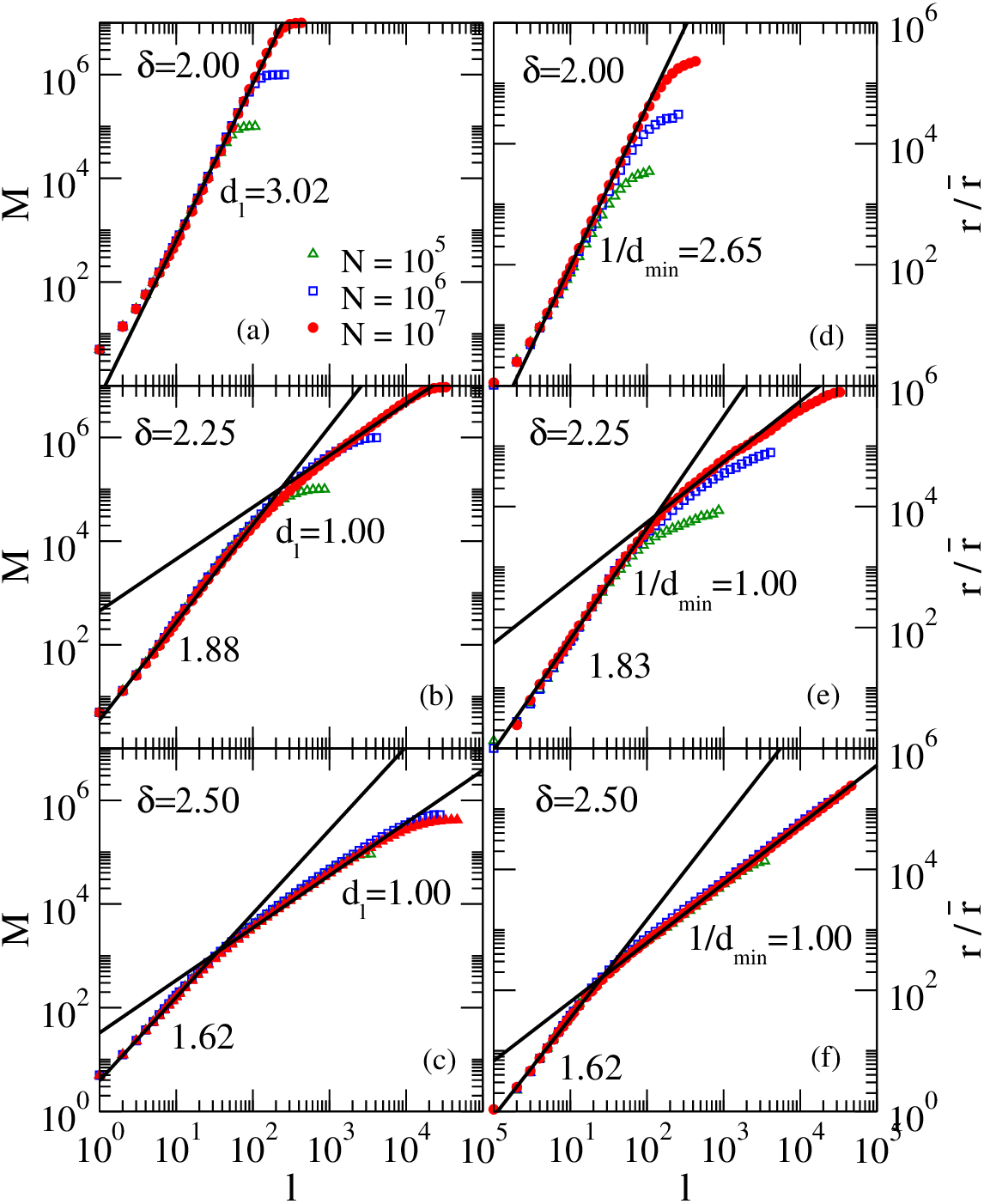}
\caption{The mass $M$ (left column) and the relative distance $r / \bar{r}$
  (right column) as a
  function of the topological distance $\ell$ for ER networks embedded in linear chains  with $k=4$,
  for the system sizes $N=10^5,\, 10^6
\,\textrm{and}\, 10^7$ with 
$\delta=2.0, \, 2.25 \,\textrm{and}\, 2.5$. 
The straight lines are best fits to the data that yield the topological dimension $d_l$ and the
dimension of the shortest path $d_{min}$. Note that the slopes below the crossover in (b), (c), (e)
and (f) of $M$ and $r$ vs $\ell$ are the same. This yields $d=1$ for all range of $r$ as indeed seen in
Fig. \ref{MR_1d}. 
}
 \label{ML2_1d}
\end{figure}
\begin{figure}[tbh!]
\centering
  \includegraphics[angle=0, scale=0.45]{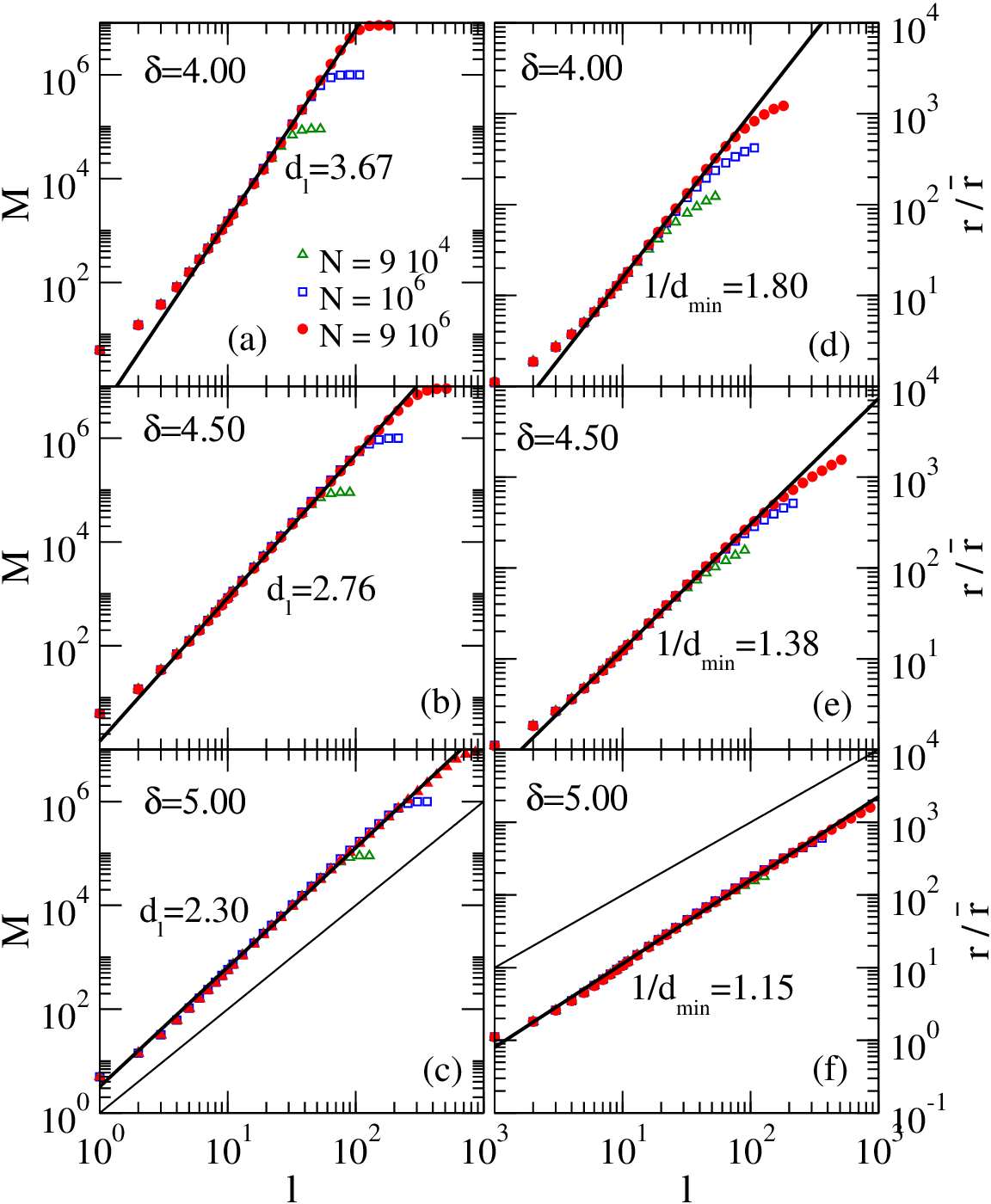}
\caption{The same as Fig \ref{ML2_1d}, but for ER networks embedded in a square lattice. The system sizes 
are $N=9 \cdot 10^4,\, 10^6 \,\textrm{and}\, 9 \cdot 10^6$ 
with $\delta=4.0, \, 4.5 \,\textrm{and}\, 5.0$. The lines in (c) and (f) demonstrate for comparison
slopes 2 and 1 respectively.
}
\label{ML2_2d} 
\end{figure}
where $d_\ell$ is the topological (''chemical'') dimension and $d_{\rm{min}}$ is the dimension
of the shortest path, see e.g., \cite{Bunde1991,havlin:Adv:Phys2002}. 
For regular lattices of dimension $d_e$, $d_\ell = d_e$ and $d_{\rm{ min}} = 1$. Thus we expect that for 
$\delta\geq 2 d_e$, 
the power law relations (8) hold. \\
\indent For $\delta=0$ the network has no spatial constraints and it is known that  the mean topological distance 
$\langle\ell\rangle$  between 2 nodes on the network scales with the network size $N$ as 
$\langle\ell\rangle\sim \log N$ \cite{Bollobas2001}. This represents the small world
nature of random graphs. Since $N$ plays the role of the mass $M$ 
of the network, it follows that  $M$ increases  exponentially with $\ell$,
i.e. $M(\ell)\sim\exp(A\ell)$. We
expect that this relation holds for $\delta < d_e$
 where $r_{\rm max}$ and $r/ \bar{r}$ are both proportional to the linear scale $L$ of the
network, see Eqs. (\ref{rbar}) and (\ref{rmax}).
Since for $\delta>2d_e$ we expect power law relations (8), we conjecture that in the
intermediate regime $d_e \leq \delta<2d_e$, $M(\ell)$ will increase
slower than exponential and faster than a power law,
via a stretched exponential,
\begin{equation}
M(\ell)\sim \exp(A\ell^\alpha), \quad d_e \leq \delta<2d_e.
\label{Mstretch}
\end{equation}
This function can bridge between the exponential behavior for $\delta<d_e$ and the power law for $\delta>2d_e$. 
For $\delta$ approaching $d_e$ from above, $\alpha$ should approach 1, while for $\delta$
approaching $2d_e$ from below, $\alpha$ should approach 0,
consistent with a power law. The conjecture, Eq. (\ref{Mstretch}) is supported by earlier numerical
simulations \cite{kosmidis2008} where it was found
that in the intermediate regime,
$\ell$ scales as $(\log N)^\beta$, leading to $\alpha=1/\beta$. On the basis of numerical
simulations it was estimated
\cite{kosmidis2008}, that $\alpha \simeq \delta (2-\delta)$ in $d_e = 1$ and
$\alpha \simeq \delta (4-\delta)/4$ in $d_e= 2$,
which actually can be combined into a single equation, $\alpha = \delta^{\prime}(2-\delta^{\prime})$, when the
relative distance exponent $\delta^{\prime}=\delta/d_e$ is introduced. Thus our conjecture (\ref{Mstretch}) becomes
\begin{equation}
 M(\ell) \sim \left\{
    \begin{array}{ll}
       e^{A\ell}    &, \delta^{\prime} <1  \\[2mm]
      e^{A\ell^{\delta^{\prime}(2-\delta^{\prime})}}     &, 1 \leq \delta^{\prime} < 2,
    \end{array}
\right.
\label{stretched}
\end{equation}
\noindent where the prefactor $A$ may depend on $\delta^{\prime}$ and $d_e$. To test this
hypothesis, we have plotted, in
Figs. \ref{ML1_1d}, a, b, c
$(d_e = 1)$ and Figs. \ref{ML1_2d} a, b, c $(d_e = 2)$, $M(\ell)$ versus
$\ell^{\delta^{\prime}(2-\delta^{\prime})}$, in a semi-logarithmic fashion. The relative distance
exponents $\delta^{\prime}$ are 0.5, 1.25 and 1.75 in both cases. The lattice sizes are the same as
in Figs. \ref{MR_1d} and \ref{MR_2d}. For $\delta^{\prime}=0.5$  where the spatial
constraints are irrelevant, we find $\log M \sim \ell$, in agreement with (\ref{stretched}).
In the intermediate $\delta$ regime $1 \leq \delta^{\prime} < 2$ we find that 
$\log M \sim \ell^\alpha$, with $\alpha$ = 0.93 ($\delta=1.25$) and 0.43 ($\delta=1.75$), also in agreement
with (\ref{stretched}). Accordingly, in the intermediate $\delta$-regime, $M(\ell)$ scales
with the topological distance $\ell$ as a stretched exponential which serves as a ''bridge'' between
the exponential behavior for $\delta < d_e$ and the anticipated power law behavior for $\delta$ well
above $2 d_e$.\\
\indent Now the question arises how the power law in Eq. (\ref{MR}) that describes the scaling of $M$ with
$r$ and the stretched exponential in
Eq. (\ref{stretched}) that describes the scaling of $M$ with $\ell$, can be
simultaneously satisfied. The only way to fulfill both equations is, that also $r(\ell)$ is a stretched exponential
with the same $\alpha$ in the intermediate regime i.e.,
\begin{equation}
r(\ell) \sim e^{B\ell^{\delta^{\prime}(2-\delta^{\prime})}}, 1 \leq \delta^{\prime} < 2,
\end{equation}
and the ratio between the prefactors $A$ and $B$ should yield  the dimension of the
network. This is since $M(\ell) \sim e^{A\ell^{\delta^{\prime}(2-\delta^{\prime})}} =
(e^{B\ell^{\delta^{\prime}(2-\delta^{\prime})}})^{A/B} \sim r^d$. Figs. \ref{ML1_1d} e, f and
\ref{ML1_2d} e, f support the assumption (11). The prefactor $B$ is obtained from
the slopes of the straight lines in the figures and indeed the values of $A/B$ are found to be
identical to the values of
the dimensions we obtained in the previous section.   For $\delta$ below $d_e$ (see
Figs. \ref{ML1_1d}d and \ref{ML1_2d}d), $r$ is independent of $\ell$ and $ M \sim e^{A\ell}$ 
(see Figs. \ref{ML1_1d}a and \ref{ML1_2d}a).
\\
\indent For $\delta \geq 2 d_e$, we expect that $M(\ell)$ and $r(\ell)$ follow power laws, such
that we can determine, from a double logarithmic plot, the chemical dimension $d_\ell$ and the
dimension of the shortest path, $d_{min}$. 
Figures \ref{ML2_1d}  and \ref{ML2_2d} show that this is the case. But surprizingly, for $\delta \geq 2d_e$ (but close to 2$d_e$), the values of $d_{min}$ and
$d_{\ell}$ do not agree  with the values for the corresponding regular
lattices. 
For $\delta = 2 d_e$, we obtain $d_\ell \simeq 3.02$ in $d_e = 1$ and $d_\ell \simeq 3.67$ in $d_e =
2$, significantly higher than the corresponding values $d_\ell=1$  and $d_\ell=2$ in regular lattices. 
Furthermore, the dimension of the shortest path $d_{min}$ is
considerably smaller than in regular lattices ($d_{min}=1$), $d_{min}=1/2.65=0.38$ in $d_e=1$ and 
$d_{min}=1/1.80=0.56$ in $d_e=2$. Since
$M \sim \ell^{d_\ell} \sim r^{d_{min}d_\ell}$, the dimension $d$ of the network for $\delta \geq 2
d_e$ is simply $d = d_{min} d_\ell$, which yields $d \simeq 1.14$ in $d_e = 1$ and $d \simeq 2.04$ in $d_e =
2$, in agreement with our results of Figs. \ref{MR_1d} - \ref{P0_1d2d}. 
For $\delta$ above $2d_e$ we expect that $d_\ell$ and $d_{min}$ accept the values of the
corresponding regular lattices. 
Figure \ref{ML2_1d} shows that this is indeed the case in $d_e = 1$, with
a pronounced crossover behavior for $\delta = 2.25$ and 2.5. The crossover point decreases with
increasing $\delta$. In $d_e = 2$, in contrast, for $\delta = 2.25d_e$  and $2.5d_e$ the
dimensions do not seem to reach their anticipated values $d_e = 2$ and $d_{min} = 1$, even though $d \cong 2$ was
obtained 
for both $\delta$ values.  Figure \ref{ML2_2d} does not suggest that this is a finite size effect since a
bending down for larger system sizes cannot be seen similar to that in $d_e=1$. However, we cannot
exclude the possibility that at
very large system sizes that right now cannot be analyzed  with the current state-of-the-art computers, there
will be a crossover towards the anticipated values of  $d_\ell=2$ and $d_{min}=1$.
\section{Summary}
\label{summary}
In summary, we studied the effect of spatial constraints on complex networks where 
 the length $r$ of each link was taken from a power law distribution, Eq. (\ref{prD}), characterized
 by the exponent $\delta$. Spatial constraints are
 relevant in all networks where distance matters, such as
 the Internet, power grid networks, and transportation networks, as well as in cellular phone
 networks and collaboration networks
 \cite{Albert2002,Satorras2004:Cambridge,Barth2011:PhysRep, Hao2009,Nowell2005, Lambiotte2008}.
Our results suggest
that for $\delta$ below the embedding dimension $d_e$, the dimension of the network is infinite  as
in the case of netwoks that are not embedded in space (represented by $\delta=0$).
For $\delta$ between $d_e$ and 2$d_e$, the dimension decreases monotonically, from $d=\infty$ to
$d=d_e$. Above 2$d_e$, $d=d_e$. We also studied how
the mass $M$ and the Euclidean distance $r$ scale
with the topological distance $\ell$. For $\delta$ below $d_e$, $M$ increases exponentially
with $\ell$, while $r$ does not depend on $\ell$. For
$\delta$ between $d_e$ and 2$d_e$, both the mass $M$ and the Euclidean distance $r$
increase with $\ell$ as a stretched exponential, with the same exponent $\alpha$ but different
prefactors in the exponential. The ratio between these two prefactors yields the dimension of the
embedded network. 
Exactly at $\delta=2d_e$, the exponent $\alpha$ becomes zero and $M$ and $r$ scale with $\ell$ as 
 power laws, defining the exponents $d_\ell$ and $d_{\rm{min}}$, respectively similar to fractal 
structures \cite{Bunde1991,havlin:Adv:Phys2002}. While the dimension
$d$ is equal to $d_e$, surprisingly 
 $d_\ell$ and $d_{\rm{min}}$ do not have the values  $d_\ell=d_e$ and $d_{\rm{min}}=1$ that are
expected for regular lattices. This effect seems to hold
in $d_e=2$ also for $\delta$ values somewhat greater than 2$d_e$.\\
\indent Our results have been obtained for a nearly $\delta$-functional degree distribution, but we
argue that they are valid for any narrow degree distribution,
like Possonian, Gaussian or exponential degree distribution since all those networks are expected to
be in the same universality class. For power law degree distributions
(scale free networks \cite{Barabasi1999}), there may be differences for small
values of $\delta$, since it is known that nonembedded random graphs and scale free networks are in
different universality classes \cite{Cohen2002:PhysRevE,Dorogovtsev2008}.  In the relevant
intermediate $\delta$ regime ($d_e \leq \delta < 2d_e$), we cannot exclude the
possibility that the dimensions do not depend on the degree distribution. Indications are from
measurements of the dimension of the airline network and
the Internet \cite{daqingDim2011}. Both are scale free networks, with
$\delta$ close to 3 (airline network) and $\delta$ close to 2.6 (Internet). For the airline network,
$d$ is close to 3, while for the Internet, $d$ is
close to 4.5. These values are consistent with those obtained here for
the ER-networks, with the same $\delta$-values.
We have assumed a power law distribution, Eq. (\ref{prD}), for the link length. Other distributions
are possible, for example an exponential distribution
which holds for the power grid and ground transportation networks \cite{Barth2011:PhysRep}. This
case is equivalent to
$\delta=\infty$, since we have a finite length scale and
thus the dimension $d$ of the network is expected to be the same as the dimension of the embedding
space $d_e$.
\\
\indent A power law distribution of Euclidean distances appears also in  other physical
systems where the present results may be relevant. For example,  model systems where the interactions between particles decay as $r^{-\delta}$ have been
  studied extensively for many years, for recent reviews on the statistical physics and dynamical
properties  of these systems, see \cite{Mukamel2008,Campa2009}.
 Magnetic models on lattices with long range bonds whose lengths follow a power law distribution  have  also  been studied, see
 e.g., \cite{Chang2007}. 
In Levy flights and walks, the jump lengths follow a power law distribution. For reviews see
\cite{Klafter1996,Metzler2000, Klafter2011}.
Finally, it has been found that a power law  distribution of link lengths with
 $\delta=d_e$ or $d_e+1$ (depending on the type of transport) is optimal for navigation 
\cite{Viswanathan1999,Kleinberg2000,Li2010,Roberson2006,Yanqing2011}.

\ack
AB and SH are grateful to the Deutsche Forschungsgemeinschaft for financial support. DL thanks the National Science Foundation of China (NSFC) for financial support.

\end{document}